\documentclass[12pt]{article}
\usepackage[english]{babel}
\usepackage[latin1]{inputenc}
\usepackage{times}
\usepackage[T1]{fontenc}
\usepackage{epsfig}
\usepackage{amsmath}
\usepackage{amssymb}

\begin{document}

\title{Didactic derivation of the special theory of relativity  from the Klein-Gordon equation}

\author{H. Arod\'z$$    \\$\;\;$ \\ \emph{ Institute of Physics,
Jagiellonian University, Cracow, Poland }}

\date{$\;$}

\maketitle

\begin{abstract}
 We present a  didactic  derivation of the special theory of relativity  in which  Lorentz transformations are `discovered'  as symmetry transformations of Klein-Gordon equation.  The  interpretation of  Lorentz boosts as  transformations to  moving inertial reference frames is  not assumed  at the start, but it naturally appears at a later stage.   The relative velocity $\textbf{v}$ of two inertial reference frames  is defined in terms of the elements of the pertinent Lorentz matrix, and the bound  $|\textbf{v}| <c\;$   is obtained as a simple theorem that follows from the structure of the Lorentz group.   The polar  decomposition of  Lorentz matrices is used to explain  noncommutativity and nonassociativity of  the relativistic composition  (`addition') of velocities. 
\end{abstract}

\vspace*{2cm} 
%\noindent PACS:  03.50.Kk \\

\pagebreak

\section{ Introduction}
The special theory of relativity  (STR) is known for more than hundred years.  Nevertheless,  its  counter-intuitive predictions about  relativity of basic space-time relations,  such   as   simultaneity of events  or spatial  distances between them,  still  attract attention.     Also the  resulting kinematics of motion of particles  is far from obvious:   the relativistic composition law  for velocities is  nonlinear, noncommutative  and  nonassociative;  and   a massive particle at rest has  the non-vanishing energy  equal to $m_0 c^2$.  Not surprisingly,  one can find   in literature a whole  variety  of  pedagogical  presentations of  STR,  including popular ones,  or addressed to physicists,  or strictly mathematical,  as the  representative examples  see  \cite{1},  \cite{2},  \cite{3},  respectively.  Below we  propose yet another one.  Our motivation for  working it out  is twofold.  First,  we would like to  emphasize in our approach  the fact that the physical origin of  Lorentz transformations lies in symmetries of  field equations. Historically, it was the electromagnetic field in vacuum and Maxwell equations, but nowadays one should also remember about other field equations.  We  choose  Klein-Gordon equation for a scalar field as the simplest representative of them. 

The second part of our motivation is to present a compelling argument  for the  bound   $|\textbf{v}| <c\;$ on the relative velocity of  inertial reference frames \footnote{We faced this problem teaching  classical electrodynamics to undergraduate students of physics.}.  We have not found a satisfactory discussion of it  in literature  available to us.  Most often it is said that
 $|\textbf{v}| <c$  because otherwise  $\sqrt{1- \textbf{v}^2/ c^2}$  would be imaginary.  But this argument  is not  fully convincing  --  an inquiring student says  that perhaps in the case of  super-luminal velocities  we should just use another form of Lorentz transformations, e.g., with   $\sqrt{\textbf{v}^2/ c^2 -1}$.  The solution we propose is to introduce Lorentz transformations first,  without  mentioning the velocity, and to define the velocity at a later stage. 
 Furthermore,  we would like to give a precise answer also to another  question about velocities:  does STR  allow for super-luminal velocities or not?  

In our didactic proposal we start from the Lorentz group  introduced  simply as a set of coordinate transformations that  leave invariant the form of Klein-Gordon equation.  A priori no physical interpretation of these transformations is assumed.  In the second step,  their   interpretation  as  transformations between uniformly  moving  reference frames is deduced.  It is suggested by the mathematical form of the Lorentz transformations of Cartesian coordinates in space-time.  Also  the relative  velocity $\textbf{v}$ of two inertial reference frames is introduced  in this step.  It is defined in terms of elements of  the Lorentz matrix.  The  bound $|\textbf{v}| <c$  appears as  a simple mathematical consequence of the structure of the Lorentz group. Apart from fulfilling the didactic tasks mentioned above,  our approach shows  that  STR is a nice example of  theory  that provides its own physical interpretation, as opposed to, e.g.,  quantum mechanics, where various physical interpretations seem to be added by hand (and hotly debated).   This aspect can render the presented approach interesting also outside the classroom.   

The layout of this article is as follows.  In the next Section  we  introduce the Lorentz group,  establish its physical interpretation,    define the relative velocity  $\textbf{v}$ of two inertial reference frames, and prove that  $|\textbf{v}| < c$.  In  Section 3  we  show that the most general  Lorentz transformations  involve nothing  else than the  well-known boosts and rotations, apart from time and space reflections. We also digress on the mentioned above unusual properties of  the relativistic composition of velocities.  Section  4  is devoted to a brief discussion of  velocities of physical particles.  Discussion of our approach  is given in Section 5. 

A word about  our notation.  We use the Cartesian coordinates  $x^i$  in the three-dimensional space. Together with   $x^0 =c t$, where $t$ is the time,  they provide the Cartesian  coordinates $x^{\mu}$ in  the four-dimensional  manifold encompassing the space and time (the space-time).  Each concrete choice of such coordinates  is called  the  Cartesian coordinate frame in the space-time and denoted as $(x^{\mu})$.  Such a coordinate frame is associated with the appropriate  coordinate net in the space-time, in analogy to coordinate nets on geographical maps.  The Latin indices take values 1,2,3, while the Greek ones $0, 1, 2, 3$.  Summation over  repeated  indices is  understood.  Often we will  use the matrix notation. In particular, the space-time coordinates form the one-column matrix \[x= \left[\begin{array}{c} x^0\\x^1\\x^2\\x^3 \end{array} \right] = [x^{\mu}]. \]  
The boldface denotes three-element  columns  of coordinates, e.g.,  \[{\bf u}= \left[\begin{array}{c} u^1\\u^2\\u^3 \end{array} \right] = [u^{i}], \]  
and  $|{\bf u}| = \sqrt{u^i u^i}.$    The numbers $u^i$ we call the coordinates of ${\bf u}$, not components, because  we  prefer the terminology in which  component of a vector is also a vector, e.g., \[{\bf u}_1= u^1 \left[\begin{array}{c} 1\\0\\0 \end{array} \right]  \]  is the first component of   ${\bf u}$.

\section{Lorentz transformations  and  the bound $|\textbf{v}| < c$}

Of course, we have to use certain empirical input about the material world.  We think that nowadays one of the best choices is Klein-Gordon equation   
\begin{equation}
\left( \triangle - \frac{1}{c^2} \frac{\partial^2}{\partial t^2}    + \lambda_C^{-2} \right)  \psi(\textbf{x},t) =0,
\end{equation}
where $c$ = 299 792 458 m/s,   $\lambda_C  > 0$ is a constant of dimension of length called  the Compton  wave length,  and  $\triangle $  is the three-dimensional  Laplacian.  The time is measured in seconds and the distance in meters.  $c$  and  $\lambda_C $  are regarded here as  phenomenological constants. The Klein-Gordon  equation   is the  cornerstone of  field theory and particle physics if we neglect the gravitational interaction.   In particular, it appears in classical electrodynamics as the wave equation for electric and magnetic fields -- in this case $\lambda_C^{-2}  =0$. In fact,   all field equations of the Standard Model of particle physics  are closely related with  it.   Loosely speaking,   Klein-Gordon equation  encodes some of the physics of non-interacting fundamental quantum particles \footnote{ Actually,  in most cases this equation is derived from more fundamental equations like Maxwell or Dirac equations, but this is not important here.}.  The empirical evidence about the fundamental particles shows that  the constant $c$ is  universal, i.e., it is common for all  of them.  This means that  $c$  characterizes rather the space-time in which the particles live than the particles  themselves.  On the other hand,  $\lambda_C $  is not universal,   it varies from particle to particle ($\lambda_C^{-1}$  is  proportional to the rest mass of the particle).  

 Note that  in order to write Klein-Gordon equation  (1)  for the free quantum particles  we have picked a Cartesian coordinate frame in the space-time.  By definition this frame is called the inertial one.   For the correct description of  physical phenomena in non-Cartesian coordinates  such as  spherical ones or others,  or in a non-inertial coordinate frame,  Eq.  (1) would  have to be  replaced by  another one  having a different mathematical  form.  In most cases,  the constant coefficients in front of derivatives would be replaced by certain nontrivial  functions of the coordinates, also some new terms could appear on the left hand side.  

 The differential operator
\[ \Box = \triangle  - \frac{1}{c^2} \frac{\partial^2}{\partial t^2}  \]  present in (1)  can be written in the  Beltrami--Laplace form,   namely
\[ \Box = - \frac{1}{\sqrt{- \eta}} \frac{\partial}{\partial x^{\mu}} \left( \eta^{\mu\nu}  \sqrt{- \eta} \frac{\partial}{\partial x^{\nu}}\right).
\]
The coefficient $\sqrt{- \eta}$, where $\eta = \mbox{det}\: \hat{\eta}$ is trivial in our case,  $\sqrt{- \eta} = 1$.  The Beltrami--Laplace operator is well-known in mathematics,  in particular in the differential geometry of manifolds which possess a metric tensor  $\hat{g} = [g_{\mu\nu}]$.  In that case, $-\hat{\eta}$  in the $ \Box$ operator is replaced by the metric tensor  $\hat{g}$.    For us  the  important point is that  the  presence of this operator in the Klein--Gordon equation  strongly suggests that the space-time  should be endowed with the metric  $\hat{\eta}$ which in the Cartesian coordinates $x^{\mu}$ has the Minkowski form 
\[\hat{\eta} = \mbox{diag}[1,-1,-1,-1] =[\eta_{\mu\nu}]. \]
 Thus the  Minkowski metric  $\hat{\eta}$
  is the diagonal matrix with the shown entries on its diagonal. We will use also  the inverse metric tensor $\hat{\eta}^{-1} = [\eta^{\mu\nu}]$.   Here the first  index   $\mu$  enumerates the rows, and  the second index    $\nu$  the columns.  In the Cartesian coordinates   $\hat{\eta}^{-1}$ numerically coincides with $\hat{\eta}$,   $\;\hat{\eta}^{-1} = \mbox{diag}[1,-1,-1,-1]  $.  
  With  such justification,  we accept the  existence of the metric  $\hat{\eta}$ as the  fundamental  mathematical  property of the space-time,  from now on called the Minkowski space-time.

Now,  being  aware of  the presence of the metric,  let us ask about  its symmetries,  that is the coordinate transformations  that leave  its form unchanged.   The metric  always is a covariant tensor of  rank two, hence  the elements of  $\hat{\eta}$ transform  according to the following  formula    
\begin{equation}  \eta'_{\rho\sigma} \: \frac{\partial x^{'\rho}}{\partial x^{\mu}}  \frac{\partial x^{'\sigma}}{\partial x^{\nu}}  = \eta_{\mu\nu}. 
\end{equation}
The Lorentz transformations   are defined as the linear transformations  of the coordinates 
\begin{equation}
x^{'\mu} = L^{\mu}_{\;\;\nu} x^{\nu}, 
\end{equation}
that leave the Minkowski metric $\hat{\eta}=[\eta_{\mu\nu}]$ invariant \footnote{The assumption of linearity is in fact  superfluous   --  one can prove that  formula  (2) and  condition  (4)  imply  that  $x^{'\mu} = L^{\mu}_{\;\;\nu} x^{\nu} + a^{\mu}$, where   $L^{\mu}_{\;\;\nu}$ and $a^{\mu}$ are constant, i.e., they do not depend on  $x^{\mu}$,  see \cite{4}.  We have put   $a^{\mu}=0$  because we are not interested in translations  in space and time. },  
\begin{equation}  \eta'_{\rho\sigma} =  \eta_{\rho\sigma}.\end{equation} 
By assumption,   $ L^{\mu}_{\;\;\nu}$  are  real  numbers,  otherwise  formula (3) would give   rather strange, hard to interpret,  complex  time and  space coordinates  $x^{'\mu}$.   They  form the four by four matrix  $\hat{L} = [L^{\mu}_{\;\;\nu}]$  called the Lorentz matrix.  The first index, $\mu$, enumerates the rows, and the second index, $\nu$, the columns. 

Formulas  (2), (3), (4)   lead to the following condition
\begin{equation}
 \eta_{\rho\sigma}  \: L^{\rho}_{\;\;\mu}  L^{\sigma}_{\;\;\nu}   = \eta_{\mu\nu},  
\end{equation}
which is the necessary condition for  $\hat{L}$ to be the Lorentz  matrix.  One can easily see that it is also the sufficient condition.  Note that the invariance of the Minkowski metric tensor means that the new coordinates $x^{'\mu}$  also are Cartesian ones.   

 Symmetry  of the metric is a  very important  mathematical  characteristics of the  space-time.   Moreover,   it can be utilized in order to generate new solutions of the Klein-Gordon equation.  If $\psi(x^i, t)$ is a solution of this equation, then  so are  all functions of the form $\psi'(x^i, t) =\psi(L^i_{\;\;\mu}x^{\mu}, L^0_{\;\;\nu}x^{\nu}/c)$. This assertion can be checked simply by differentiation of  these new functions as required  on the l.h.s. of  equation (1) and using  identities (9) obtained below.   The closely related fact is that  the operator  $\Box$ does not  change its form when expressed by the new coordinates  $x^{'\mu}$.  Let us add that operations that transform one solution into another are called  symmetries  of the pertinent equation, here of the Klein-Gordon equation.

The matrix form of (3)   reads   $ x' = \hat{L} x$,  and  (5) is equivalent to
\begin{equation}
\hat{L}^T \hat{\eta} \:\hat{L} = \hat{\eta}. 
\end{equation}
Here   $ \hat{L}^T $ is the transposed 
matrix, i.e.,  $(\hat{L}^T)^{\mu}_{\;\;\nu} = L^{\nu}_{\;\;\mu} $.    The set of all real, 4 by 4 matrices  obeying condition (6) is a group with the multiplication  of group elements  given by the  matrix product.   It is called the full Lorentz group.  
Taking the determinant of both sides of condition  (6)  we see that  $\mbox{det}\: \hat{L} = \pm 1, $ hence $\hat{L} ^{-1}$ exists. It is easy to see that  $\hat{L} ^{-1}$  obeys that condition  too.  Note also that     $\hat{L} ^{-1} =\hat{\eta}^{-1}\hat{L}^T \hat{\eta}$.  Therefore,  
\begin{equation} 
(\hat{L} ^{-1})^{\mu}_{\;\;\nu} = \eta^{\mu\lambda} L^{\sigma}_{\;\;\lambda} \eta_{\sigma \nu} \equiv  L_{\nu}^{\;\;\mu} .
\end{equation}
The notation on the r.h.s. reflects the fact that  in the case of tensors
 $\hat{\eta}^{-1}$ raises  and $\hat{\eta}$ lowers indices. We use it because it is convenient,  even though $\hat{L}$ is  the transformation matrix and not a tensor. 

 Simple algebraic manipulations  on condition (6) give  the  equivalent form
\begin{equation}
 \hat{\eta}^{-1} =  \hat{L} \:\hat{\eta}^{-1} \:\hat{L}^T.  
\end{equation}
which  for the matrix elements  reads 
\begin{equation}
 \eta^{\mu\nu} = L^{\mu}_{\;\;\rho}  L^{\nu}_{\;\;\sigma} \: \eta^{\rho\sigma}.    
\end{equation}

Until now,  all steps  have been  purely mathematical ones.  As physicists,  we  are interested in  actual performing  transformations  (3)  in the material world.  In other words, what is the physical interpretation of  these transformations?  It turns out that the formulas we have obtained give a tantalizing hint about it. 

Let us write (3) in the expanded form, 
\begin{equation}
t' = L^0_{\;\;0} t + L^0_{\;\;k} x^k/c, \;\;\; x^{'i} =c  L^i_{\;\;0}  t  + L^i_{\;\;k} x^k. \end{equation}
Condition (9)  in the case $\mu=\nu=0$ gives  \[(L^0_{\;\;0})^2= 1+ L^0_{\;\;i}L^0_{\;\;i}. \] Because $L^0_{\;\;i}$  are real,  we see that  $|L^0_{\;\;0}| \geq 1$.  This implies  that $L^0_{\;\;0} \neq 0$, hence  we may  calculate $t$ from the first formula  (10),   \[ t = \frac{1}{L^0_{\;\;0}} \left( t' - L^0_{\;\;k} x^k/c\right). \]  Inserting it  in the second formula (10) gives
\begin{equation} x^{'i} = c \frac{L^i_{\;\;0}}{ L^0_{\;\;0}} t'+ \left( L^i_{\;\;k} -  \frac{L^i_{\;\;0} L^0_{\;\;k}}{ L^0_{\;\;0}}\right) x^k. \end{equation}
This formula shows that  a  material point  that has constant, i.e., time independent, coordinates $x^i$  moves with the velocity  
\begin{equation} 
w^{'i}= c \frac{L^i_{\;\;0}}{ L^0_{\;\;0}} 
\end{equation} 
with respect to the new coordinate frame $(x^{'k})$.   This velocity does not depend on  the position in the space  given by $\textbf{x}= [x^k]$.    Thus, $\textbf{w}' = [w^{'i}]$ is the velocity of the frame $(x^{\mu})$ with respect to $(x^{' \mu})$.  Its coordinates $w^{'i}$ are given with respect to the frame $(x^{'\mu})$ because
$ w^{'i} = d x^{'i}/dt'$.

Transformation  (10)  can easily be inverted by  replacing  $\hat{L}$  with $ \hat{L}^{-1}$.  Using  (7) we  obtain  
\begin{equation}
t = L_0^{\;\;0}\: t' + L_k^{\;\;0}\: x^{'k}/c, \;\;\; x^{i} =c  L_0^{\;\;i} \: t'  + L_i^{\;\;k} \:x^{'k}. \end{equation}
Next, using  the first formula  (13)  we may express $t'$  by $t$  in  the second formula, 
\begin{equation} x^{i} = c \frac{L_0^{\;\;i}}{ L_0^{\;\;0}} t+ \left( L_k^{\;\;i} -  \frac{L_0^{\;\;i}
 L_k^{\;\;0}}{ L_0^{\;\;0}}\right) x^{'k}. \end{equation}
We see that  the velocity of the frame  $(x^{' \mu})$ with respect to the frame $(x^{ \mu})$  is given by formula 
\begin{equation} 
v^i = c \frac{L_0^{\;\;i}}{L_0^{\;\;0}},
\end{equation}
where the coordinates $v^i$  of the velocity are  given in the frame  $(x^{ \mu})$.    We shall see in the next Section that the coordinates of the two velocities  are related to each other,  but in general  $v^i \neq - w^{'i}$.  

The definitions (12) and (15) give 
\begin{equation}
v^i  v^i = w^{'i} w^{'i} = c^2  \left(1 - \frac{1}{(L^0_{\;\;0})^2}\right).
\end{equation} 
This formula is obtained with the help of  the relations  $ L^0_{\;\;i} L^0_{\;\;i} = (L^0_{\;\;0})^2 -1$,   $\; L_0^{\;\;i} L_0^{\;\;i} = (L_0^{\;\;0})^2 -1$,  $\;L^0_{\;\;0} = L_0^{\;\;0}$.   Because $(L^0_{\;\;0})^2 \geq 1$, we see that  $|\textbf{w}'| = |\textbf{v}| < c$.   It is clear that  this  theorem 
follows  essentially from the  fundamental  formulas  (3) and  (5) .  

To summarize,  we have found that  there is the subluminal velocity  $\textbf{v}$  associated with  the Lorentz transformation  $\hat{L}$.  It is defined by formula (15), and interpreted as  the velocity of the Cartesian coordinate frame $(x^{'\mu})$ with respect to the original coordinate frame  $(x^{\mu})$. In our approach  this  fact  has been found as  the result of the analysis of Lorentz transformations (3), and not  assumed  as in other approaches.

\section{The full physical content  of  Lorentz  transformations}

In this Section we show that the general Lorentz  transformation can be composed from the following basic transformations:  boost with the velocity $\textbf{v}$,   spatial rotation,    time or space reflection.    This result  is very important  because it says how we can physically  perform  arbitrary Lorentz transformation, at least in principle.  The  following  operations  on the Cartesian coordinate net that defines the original coordinates $x^{\mu}$ are allowed, and  no others: setting  it in a uniform motion with the velocity $\textbf{v}$ such that  $|\textbf{v}| < c$,    rotating  the coordinate net around an axis  through the origin,    flipping directions  of  the spatial coordinate axises  to the opposite ones, and  replacing the clock with  another one with hands rotating   anti-clockwise.  

 As the first step,  we just  solve  (16)  for $L^0_{\;\;0}$.   This gives 
\begin{equation}
L^0_{\;\;0} = \pm \gamma, 
\end{equation}
where   $\gamma = (1- \textbf{v}^{\:2}/c^2)^{-1/2}$.  Now  formula (15)  yields 
\begin{equation}
L^0_{\;\;i} = \mp \gamma \frac{v^i}{c},  
\end{equation}
where we have raised the  index $0$ and lowered $i$  using $\eta^{\mu\nu}$ and  $\eta_{\mu\nu}$.

A bit longer calculation is needed in order to find the elements $L^i_{\;\;k}$ of the Lorentz matrix.  
 Condition (5)  with $\mu=i,  \:\nu = k$  reads
\[ L^0_{\;\;i}  L^0_{\;\;k}  -  L^s_{\;\;i}  L^s_{\;\;k}  = - \delta_{ik}, 
\]
where $\delta_{ik}$ is the three-dimensional Kronecker delta.   Using (18)  we obtain  the set of  quadratic equations for  the matrix elements $ L^s_{\;\;i} $:
\begin{equation}
 L^s_{\;\;i}  L^s_{\;\;k}  = \delta_{ik} + \gamma^2 \frac{v^i v^k}{c^2}.
\end{equation}
Here  the matrix algebra is helpful.  Introducing  the three by three matrix $\hat{l} =[L^i_{\;\;s}]$,  and the matrix $\textbf{v}\otimes \textbf{v} = [ v^i v^k]$ called the tensor product of  ${\bf v}$ with itself,  we write equation (19)  in the matrix form,  
\begin{equation}
\hat{l}^T\; \hat{l} = I_3 + \gamma^2  \frac{\textbf{v}\otimes \textbf{v}}{c^2}, 
\end{equation}
where  $I_3$ is the  3 by 3 unit matrix.  It is clear that  if  $\hat{l}$  is a solution of this equation then so are all matrices of the form $ \hat{R} \hat{l}$, where  $\hat{R}$  can be  arbitrary  orthogonal 3 by 3 matrix.  Such matrix  $\hat{R}$  cancels out on the l.h.s. of Eq.\ (20)  because 
$\hat{R}^T \hat{R} = I_3$.  It represents a spatial rotation around a certain axis  through the origin,  possibly accompanied  by  the spatial reflection.  Actually,  any  three by three matrix $\hat{l}$  can be written in the form 
\begin{equation} \hat{l} = \hat{R}\:\hat{h}, \end{equation}
where  $\hat{h}$ is  a   symmetric, three by three real matrix  with  non-negative eigenvalues.  
Formula (21)  gives the so called   polar decomposition  of   $\hat{l}$,   see, e.g.,  \cite{5}.   Such decomposition, usually  introduced for complex matrices,  generalizes to matrices the  well-known  decomposition of a complex number $z$ into the product of modulus and the phase factor, $z = \exp(i \:\mbox{Arg}\: z)\:|z| $.  The matrix $\hat{h}$  is the counterpart of $|z|$.  For us  the main advantage of the polar decomposition  is that 
the matrix equation (20)  is  reduced to the equation 
\begin{equation}
\hat{h}^2 = I_3 + \gamma^2  \frac{\textbf{v}\otimes \textbf{v}}{c^2}, 
\end{equation}
which has  a unique solution  for $\hat{h}$  in the indicated class of matrices.  Leaving aside direct solving the corresponding set of quadratic equations for the matrix elements of $\hat{h}$,  let us  guess  that $\hat{h}$      has  the form 
\begin{equation}
\hat{h} = \alpha I_3 + \beta \frac{\textbf{v}\otimes \textbf{v}}{c^2},
\end{equation}
where  $\alpha, \beta$ are  constants.   This  presupposition  comes from the observation that  
\[(\frac{\textbf{v}\otimes \textbf{v}}{c^2})^2 = (1- \frac{1}{\gamma^2})  \frac{\textbf{v}\otimes \textbf{v}}{c^2}. \]
Inserting the matrix (23) in Eq.\ (22) and equating the coefficients in front of the matrices  $I_3$ and  $\textbf{v}\otimes \textbf{v}/ c^2$  we  obtain quadratic equations for  $\alpha, \beta$ which are easy to solve. Taking care of signs in order to ensure that $\hat{h}$ has  non-negative eigenvalues,  we finally obtain 
\begin{equation}
\hat{h} =  I_3 +\frac{\gamma^2}{ 1+ \gamma} \:\frac{\textbf{v}\otimes \textbf{v}}{c^2},  \;\;\;\;\;\; h^i_{\;\;k} =  \delta_{ik} + \frac{\gamma^2}{ 1+ \gamma} \:\frac{v^i v^k }{c^2}. 
\end{equation}

Eigenvectors of  the matrix   (23)  are very simple. If $\textbf{v} \neq 0$, it is just $\textbf{v}$ with the corresponding eigenvalue  $\alpha + (1- \gamma^{-2}) \beta $, and two vectors orthogonal to $\textbf{v}$ and to each other  with the double degenerate eigenvalue equal to $\alpha$.  If  $\textbf{v} =0$,  we may take as the eigenvectors any three 
orthonormal vectors, and  we have the triple degenerate eigenvalue equal to $\alpha$.  The matrix 
(24)  has  the eigenvalues  $1, 1$, and  $ \gamma$.  

It remains to determine the elements  $L^i_{\;\;0}$.  Condition (9)  with  $\mu =i, \: \nu =0$ 
gives  \[L^i_{\;\;0}= L^i_{\;\;k} L^0_{\;\;k}/L^0_{\;\;0}. \]  All matrix elements  on the r.h.s.  of this formula are already known, see  (17), (18), (21) and (24).  Using these formulas we obtain
\begin{equation}
L^i_{\;\;0} = - R^i_{\;\;p} h^p_{\;\;k}\frac{ v^k}{c} = - \gamma R^i_{\;\;p} \frac{ v^p}{c}.
\end{equation} 
The second equality follows from the fact that  $\textbf{v}$ is the eigenvector of  $\hat{h}$  with the  eigenvalue  $\gamma$.  

Note that 
\begin{equation} w^{'i} = \mp R^i_{\;\;k} v^k, 
\end{equation} 
as follows from formulas (12),  (17) and (25).   This  relation between the coordinates of the relative  velocities  reflects the fact that in general  the two  coordinate nets are rotated  with respect to each other.

Our findings  are summarized by presenting  the matrix $\hat{L}$  as the  product  of matrices corresponding  to the time reflection, rotation and the spatial reflection, and pure boost:    
\begin{equation}
\hat{L} = \left[ \begin{array}{c|c}  \pm 1 & \;\;\;0\;\;\;\\ \hline\\  0 &  I_3\\ \\ \end{array} \right] \; \left[ \begin{array}{c|c}  1 & \;\;\;0\;\;\; \\ \hline\\ 0 & \hat{R} \\  &\end{array} \right] \; \left[ \begin{array}{c|c}  \gamma & \;- \gamma \frac{v^k}{c}\; \\ \hline\\ - \gamma \frac{v^i}{c} & h^i_{\;\;k} \\  &\end{array} \right].
\end{equation} 

Formula (27) essentially gives the polar decomposition of the Lorentz matrix $\hat{L}$:  it can be rewritten in the form 
\begin{equation}   
\hat{L} = \hat{{\cal O}} \hat{H}(\bf{v}),
\end{equation}
where  the matrix
\begin{equation}
\hat{{\cal O}} =  \left[ \begin{array}{c|c}  \pm 1 & \;\;\;0\;\;\; \\ \hline\\ 0 & \hat{R} \\  &  \end{array} \right]
\end{equation} 
is  simultaneously orthogonal, $\hat{{\cal O}}^T \hat{{\cal O}} = I_4$,  and Lorentz,   $\hat{{\cal O}}^T \hat{\eta}\hat{{\cal O}} = \hat{\eta}$.   It turns out that  (29) gives the most general form of  such matrices.  The matrix  
\begin{equation}
\hat{H}(\bf{v})  =  \left[ \begin{array}{c|c}  \gamma & \;- \gamma \frac{v^k}{c}\; \\ \hline\\ - \gamma \frac{v^i}{c} & h^i_{\;\;k} \\  &\end{array} \right],
\end{equation} 
often called the Lorentz boost, 
is the  symmetric Lorentz matrix with the positive eigenvalues (which are equal to  $\gamma (1 \pm |\bf{v}|/c), 1, 1$), as required in the polar decomposition.  Note that  $\hat{H}(\bf{v})  = \hat{H}(-\bf{v})$. 

The polar decomposition (28)  can be used  to explain the  generally noncommutative and nonassociative 
character of  the relativistic composition of velocities mentioned in the Introduction.  This composition law can be obtained from the formula 
\begin{equation}   
\hat{H}(\bf{v}) \; \hat{H}(\bf{u}) = \hat{{\cal O}}(\bf{v}, \bf{u})  \hat{H}(\bf{w}(\bf{v},\bf{u})). 
\end{equation}
Calculating the upper row of the matrix $\hat{H}(\bf{v}) \; \hat{H}(\bf{u})$  we find 
\begin{equation}
\gamma(\bf{w}(\bf{v},\bf{u})) = (1+ \frac{\bf{v}\bf{u}}{c^2}) \:\gamma(\bf{v}) \:\gamma(\bf{u}), 
\end{equation}
and 
\begin{equation}
\bf{w}(\bf{v},\bf{u}) = \frac{1}{1+ \bf{v}\bf{u}/c^2}\: \left[\frac{1}{\gamma(\bf{u})} \bf{v} +   \left(1+   \frac{\gamma(\bf{u})}{1+\gamma(\bf{u})} \:\frac{\bf{v}\bf{u}}{c^2} \right) \bf{u} \right]. 
\end{equation}
Let us introduce the notation  
\[ \bf{w}(\bf{v},\bf{u}) \equiv \bf{v} \vdash \bf{u}.\]
In general,  \\(a)    $\; \bf{v} \vdash \bf{u} \neq \bf{u} \vdash\bf{v}$  \hspace*{0.4cm} (the noncommutativity), \\ and \\(b)   $\; (\bf{v}_1 \vdash \bf{v}_2) \vdash \bf{v}_3 \neq \bf{v}_1 \vdash (\bf{v}_2 \vdash \bf{v}_3) $  \hspace*{0.4cm} (the nonassociativity). \\
Obviously,  it is rather inappropriate to refer to such a composition law as to the  `addition' of velocities, as it sometimes happens in literature.  

The noncommutativity follows from the fact that 
\[ \hat{H}(\bf{v}) \; \hat{H}(\bf{u}) \neq \hat{H}(\bf{u}) \; \hat{H}(\bf{v}), \] 
unless $\bf{v}$ is parallel to  $\bf{u}$  --- only in this special case  $\bf{v} \vdash \bf{u} = \bf{u} \vdash \bf{v}$. 

In order to see the nonassociativity, let us consider the identity  
\[ \left( \hat{H}(\bf{v}_1)   \hat{H}(\bf{v}_2)\right)  \hat{H}(\bf{v}_3)   =  \hat{H}(\bf{v}_1)  \left(\hat{H}(\bf{v}_2)   \hat{H}(\bf{v}_3)\right)  \]
(the matrix product is associative).  Applying formula (31)  we obtain  \[ \hat{{\cal O}}(\bf{v}_1, \bf{v}_2) \hat{H} (\bf{v}_1 \vdash \bf{v}_2) \hat{H}(\bf{v}_3) =  \hat{H}(\bf{v}_1) \hat{{\cal O}}(\bf{v}_2, \bf{v}_3)   \hat{H} (\bf{v}_2 \vdash \bf{v}_3), \]
and \[ \hat{{\cal O}}(\bf{v}_1, \bf{v}_2)  \hat{{\cal O}}(\bf{v}_1\vdash  \bf{v}_2, \bf{v}_3)   \hat{H} ((\bf{v}_1\vdash \bf{v}_2) \vdash \bf{v}_3)  \hspace*{6cm} \]   \begin{equation} \hspace*{1cm} =\hat{H}(\bf{v}_1)  \hat{{\cal O}}(\bf{v}_2, \bf{v}_3)    \hat{H}^{-1}(\bf{v}_1) \hat{{\cal O}}(\bf{v}_1,  \bf{v}_2 \vdash  \bf{v}_3) \hat{H} (\bf{v}_1 \vdash (\bf{v}_2 \vdash \bf{v}_3)).  \end{equation}
The l.h.s. of this  formula  gives the polar decomposition of $ \hat{H}(\bf{v}_1)   \hat{H}(\bf{v}_2) \hat{H}(\bf{v}_3) $.  Now, the point is that  the Lorentz matrix  $\hat{H}(\bf{v}_1)  \hat{{\cal O}}(\bf{v}_2, \bf{v}_3)    \hat{H}^{-1}(\bf{v}_1)$ present on the r.h.s. of formula (34) in general is not orthogonal.  This can be seen by checking whether it has the form (29).  If it was orthogonal, the r.h.s. of formula  (34) would also constitute 
the polar decomposition of  $ \hat{H}(\bf{v}_1)   \hat{H}(\bf{v}_2) \hat{H}(\bf{v}_3) $,  and the uniqueness of the  decomposition would imply the associativity.  It is clear that the reason for the nonassociativity  is the presence of the matrix  $\hat{{\cal O}}(\bf{v}_2, \bf{v}_3) $ in the polar decomposition of the matrix  $\hat{H}(\bf{v}_2)   \hat{H}(\bf{v}_3)$ or, in other words, the fact that superposition of two Lorentz boosts  is not a Lorentz boost,  in general.

\section{The velocities of material  particles}

The   velocity $\textbf{v}$  characterizes   the  relation of the two  Cartesian coordinate systems in the Minkowski space-time,   established by formula (3)  with $L^{\mu}_{\;\;\nu}$  obeying  the condition (5).  Velocities of material  particles are introduced in  a different  way.  The main notion here is  that of trajectory. For example,  in the case of a single point particle,  $\textbf{x}(t) = [x^i(t)]$  in the coordinate frame  $(x^{\mu})$.  In order to avoid mathematical problems with differentiation  let us consider only smooth trajectories. The  velocity $\textbf{u}(t)$ of the particle  is then defined as the derivative 
\[ \textbf{u}(t) = \frac{d \textbf{x}(t)}{d t}.  \] 
This definition does not imply any restriction on the magnitude of $|\textbf{u}|$.  If there is a restriction, it comes from the physical nature of the particle as observed  in experiments.  A priori  each of the   three  possibilities:  $|\textbf{u}| <c, \;|\textbf{u}|=c, $  $\; |\textbf{u}|>c$  is allowed for.  We know that there exist particles which  move with  the velocity $|\textbf{u}|<c$. There  also exist particles  with   $|\textbf{u}|=c$,  namely  the photons.  Perhaps certain neutrinos   also belong to this latter class.  Superluminal  particles,  called tachyons, have not been discovered yet, but there is an extensive theoretical literature on them showing that  their existence  would not generate dramatic problems,  at least  on the theory side. 

The  use of $c$ as the reference velocity in that classification is not accidental.  Precisely with this choice  the type of the particle does not  depend on  the  choice of the Cartesian coordinate  frame in the Minkowski space-time.  In the new  coordinates $x^{'\mu}$, related to the previous ones by formula (3),  the trajectory of the particle is represented by the functions $x^{'i}(t')$, and  coordinates of  the 
velocity are given by    $ u^{'i}(t')= d x^{'i}(t')/ dt'$.  Simple calculation,  in which we use formulas  (10), gives the following relation 
\begin{equation}
 u^{'i}(t')= c \:\frac{c L^i_{\;\;0} + L^i_{\;\;k} u^k(t)}{c L^0_{\;\;0} + L^0_{\;\;s} u^s(t)}. 
\end{equation}
In consequence,  
\begin{equation}
\textbf{u}^{\:'2} -c^2  =   \frac{c^2}{ (c L^0_{\;\;0} + L^0_{\;\;s} u^s)^2} (\textbf{u}^{\:2} - c^2).
\end{equation}
In the derivation of  this  formula we have used the  identities 
\[  L^i_{\;\;k}   L^i_{\;\;s} =  L^0_{\;\;k} L^0_{\;\;s}   + \delta_{ks}, \;\;  L^i_{\;\;0}   L^i_{\;\;s} =  L^0_{\;\;0}  L^0_{\;\;s},\;\; L^i_{\;\;0}  L^i_{\;\;0}  = 1+ L^0_{\;\;0} L^0_{\;\;0}  \]
 that follow directly from  condition (5).   It is clear  from (36)  that  the differences $\textbf{u}^{\:2} - c^2$ and  $\textbf{u}^{\:'2} - c^2$  always  have the same sign.

\section{Discussion}

 The presented  derivation of  Lorentz transformations  exploits the fact that they are tightly connected with  symmetries of the class of wave equations,  known as  the relativistic wave equations and represented here by the Klein-Gordon equation,  that describe matter on  the most fundamental level. The wave equations for the electric and magnetic fields of classical electrodynamics belong to this class, but also many others, for example, Dirac  and Proca equations for particles of spin 1/2 or 1, respectively.    The  Klein-Gordon equation,  its form and symmetries,  is  regarded  here as  the  initial  phenomenological input from which we start.  Such definition of the Lorentz  transformations has the rather interesting aspect that  it does not rely   on the popular in literature,  and  artificial in our feeling,   picture of  two  observers moving with a constant relative velocity.  After all,  one can think of a world with just one observer which for some reason  uses all  Cartesian coordinate systems allowed by the symmetry of Klein-Gordon equation.  Moreover, in that picture there is no a priori reason to assume that that velocity can not be arbitrarily large. 
Of course,  the uniformly moving coordinate frames  appear also in our approach, but only  at a later stage,  when we ask how to physically  accomplish the symmetry transformations.   In this sense, we emphasize the symmetry aspect,  while the relativity aspect of Lorentz transformations is deduced, similarly as  the bound  $|\textbf{v}| < c$.

Another point we would like to  emphasize is that we do not identify the constant $c$   with the velocity of light in vacuum.  It is  introduced rather as the fundamental constant of Nature, common for all fundamental particles, including massive ones.  The reason  is that   STR would be valid  also in a theoretical world  without  the photons  and  electromagnetic field, and   $c$ still  would play the  fundamental role  in such a material world.   We think that  the fact that  group velocity of packets of electromagnetic waves   in vacuum is equal to $c$  should  better be derived at a certain  stage later,  as the physical prediction from  Maxwell equations.    For this reason we are not satisfied with   presentations of STR  that start  from  the axiom that  the velocity of  light in vacuum should be the same in all 
inertial reference frames.  Such  approaches put too  strong emphasis on   the electromagnetic field.  Similar viewpoint was presented already in \cite{6}.  The choice of Klein-Gordon equation, the proxy for field equations of the Standard Model of particle physics,  as the starting point  seems more natural.  It is strongly supported by the tremendous phenomenological  success of the Standard Model  in the last  decades, with its plethora of  fundamental fields and particles.

 Our introduction to Lorentz transformations  surely does not belong to popular physics. Nevertheless,  it is rather simple -- we  mainly  use  elementary properties of  matrices.  Especially helpful is the polar decomposition, with  which we have provided the simple explanation of the peculiar properties of the composition of velocities. Moreover,  we think that  the presented approach has  also  other merits,  such as conceptual economy,  naturalness coming from being  based  directly on  the relativistic field equations, and  the clear  logical status of the  bound    $|\textbf{v}| < c$.

\section{Acknowledgement}  We thank  L. M. Sokolowski for a very informative discussion and for reading the manuscript.

\end{document}